\documentclass[Physsubmission, Phys]{SciPost}

\hypersetup{
    colorlinks,
    linkcolor={red!50!black},
    citecolor={blue!50!black},
    urlcolor={blue!80!black}
}

\usepackage[bitstream-charter]{mathdesign}
\DeclareSymbolFont{usualmathcal}{OMS}{cmsy}{m}{n}
\DeclareSymbolFontAlphabet{\mathcal}{usualmathcal}

\begin{document}

{\hfill FERMILAB-CONF-21-377-QIS-SCD-T}

\begin{center}{\Large \textbf{
DIS physics at the EIC and LHeC and connections to the future LHC and $\nu$A programs\\
}}\end{center}

\begin{center}
T.~J.~Hobbs\textsuperscript{1,2,3,4,*}
\end{center}

\begin{center}
{\bf 1} Fermi National Accelerator Laboratory, Batavia, IL 60510, USA
\\
{\bf 2} Department of Physics, Illinois Institute of Technology, Chicago, IL 60616, USA
\\
{\bf 3} EIC Center, Jefferson Lab, Newport News, VA 23606, USA
\\
{\bf 4} Department of Physics, Southern Methodist University, Dallas, TX 75275, USA
\\
* thobbs@fnal.gov
\end{center}

\begin{center}
\today
\end{center}


\definecolor{palegray}{gray}{0.95}
\begin{center}
\colorbox{palegray}{
  \begin{tabular}{rr}
  \begin{minipage}{0.1\textwidth}
    \includegraphics[width=22mm]{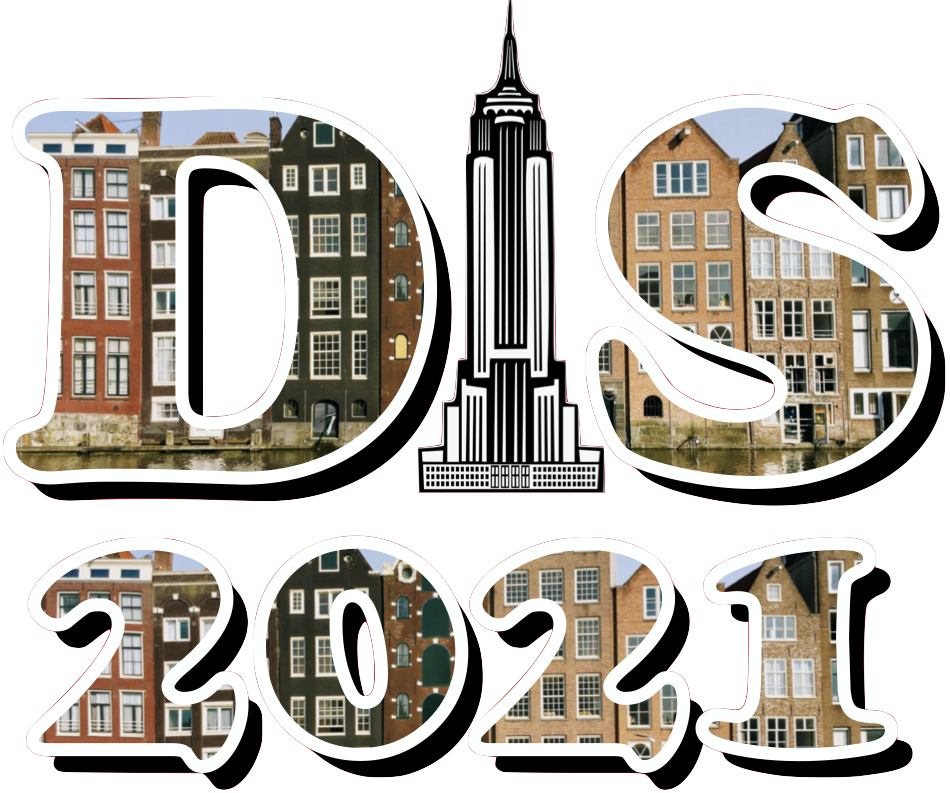}
  \end{minipage}
  &
  \begin{minipage}{0.75\textwidth}
    \begin{center}
    {\it Proceedings for the XXVIII International Workshop\\ on Deep-Inelastic Scattering and
Related Subjects,}\\
    {\it Stony Brook University, New York, USA, 12-16 April 2021} \\
    \doi{10.21468/SciPostPhysProc.?}\\
    \end{center}
  \end{minipage}
\end{tabular}
}
\end{center}

\section*{Abstract}
{\bf
Deeply-inelastic scattering (DIS) stands to enter a golden age with the prospect of precision
programs at the Electron-Ion Collider (EIC) and Large Hadron-electron Collider (LHeC). While these programs
will be of considerable importance to resolving longstanding issues in (non)perturbative QCD as well as hadronic
and nuclear structure, they will also have valuable implications for a wider range of physics at the Energy and Intensity Frontiers,
including at the High-Luminosity LHC (HL-LHC) and future $\nu A$ facilities. In this plenary contribution, we highlight a number of
salient examples of the potential HEP impact from the complementary EIC and LHeC programs drawn from their respective
Yellow Report and Whitepaper. Interested readers are encouraged to consult the extensive studies and literature from
which these examples are taken for more detail.
}

\vspace{10pt}
\noindent\rule{\textwidth}{1pt}
\tableofcontents\thispagestyle{fancy}
\noindent\rule{\textwidth}{1pt}
\vspace{10pt}

\section{Introduction}
\label{sec:intro}

The past several years have witnessed rapid development in the areas of hadronic structure and QCD as well as in efforts to test the
standard model (SM) of particle physics at both the Energy and Intensity Frontiers. Rather than being isolated one from the other,
deep complementarities connect the progress being made on these various fronts. In particular, a crucial link among the activities in these
areas is the central role of deeply-inelastic scattering (DIS) as a sensitive probe of the internal structure of hadrons and nuclei ---
a fact which follows mainly from the experimental `cleanliness' of the DIS process, as well as its ability to furnish a kinematical
lever-arm by measuring structure functions or DIS cross sections over diverse scales to constrain DGLAP scaling violations. Additionally,
theoretical control over the pure DIS process has achieved a high level of theoretical accuracy, with NNLO and, increasingly, N$^3$LO
the state-of-the-art in determining Wilson coefficients and related quantities. Critically, DIS provides direct access to the parton
distribution functions (PDFs) of the nucleon [and analogous nuclear PDFs (nPDFs) for nuclear targets]. The PDFs are an essential
nonperturbative input required for hadronic collider experiments. For instance, for the inclusive hadroproduction, $pp \to W/Z + X$,
of electroweak (EW) bosons, the ability to predict SM contributions to the cross section follows from knowledge of the perturbatively
calculable parton-level cross section, $\hat{\sigma}$, and the PDFs of the colliding protons:
\begin{equation}
	\sigma(AB\to W/Z\!+\!X)\ =\ \sum_{n}\,\alpha_{s}^{n}\, \sum_{a,b}\int dx_{a}dx_{b}\,
 f_{a/A}(x_{a},\mu^{2})\,\hat{\sigma}_{ab\to W/Z+X}^{(n)}\big(\hat{s},\,\mu^{2}\big)\,f_{b/B}(x_{b},\mu^{2})\ .
\label{eq:LHC}
\end{equation}
Although precision in tests of the SM is potentially limited by an array of experimental systematic as well as
theoretical uncertainties, PDF uncertainties are likely to increasingly dominate the landscape of error sources. This
logic applies to Higgs-production cross sections, $W$-mass determinations, extractions of $\sin^2\theta_W$, and
a multitude of searches --- direct and indirect --- for beyond SM (BSM) physics at the LHC.
Conversely, extending to lower energies, knowledge of the PDFs and related quantities are an important limitation
in Intensity Frontier activities entailing searches for a potential CP-violating phase, $\delta_\mathrm{CP}$, at
long-baseline neutrino facilities. Such efforts require detailed knowledge of the neutrino-nuclear interactions in
the few-GeV regime, including $\nu$A DIS. Here, control over the PDFs and power-suppressed corrections of
relevance at lower $Q^2$ and $W^2$ is a primary limitation in the realization of the required precision.
As a result, high-quality DIS information will play a valuable role in extending the general precision and sensitivity
to BSM physics at both the HL-LHC~\cite{ApollinariG.:2017ojx} and future $\nu A$ facilities like DUNE~\cite{1512.06148}.

\begin{figure}[h]
\centering
\vspace*{-1cm}\includegraphics[width=0.46\textwidth]{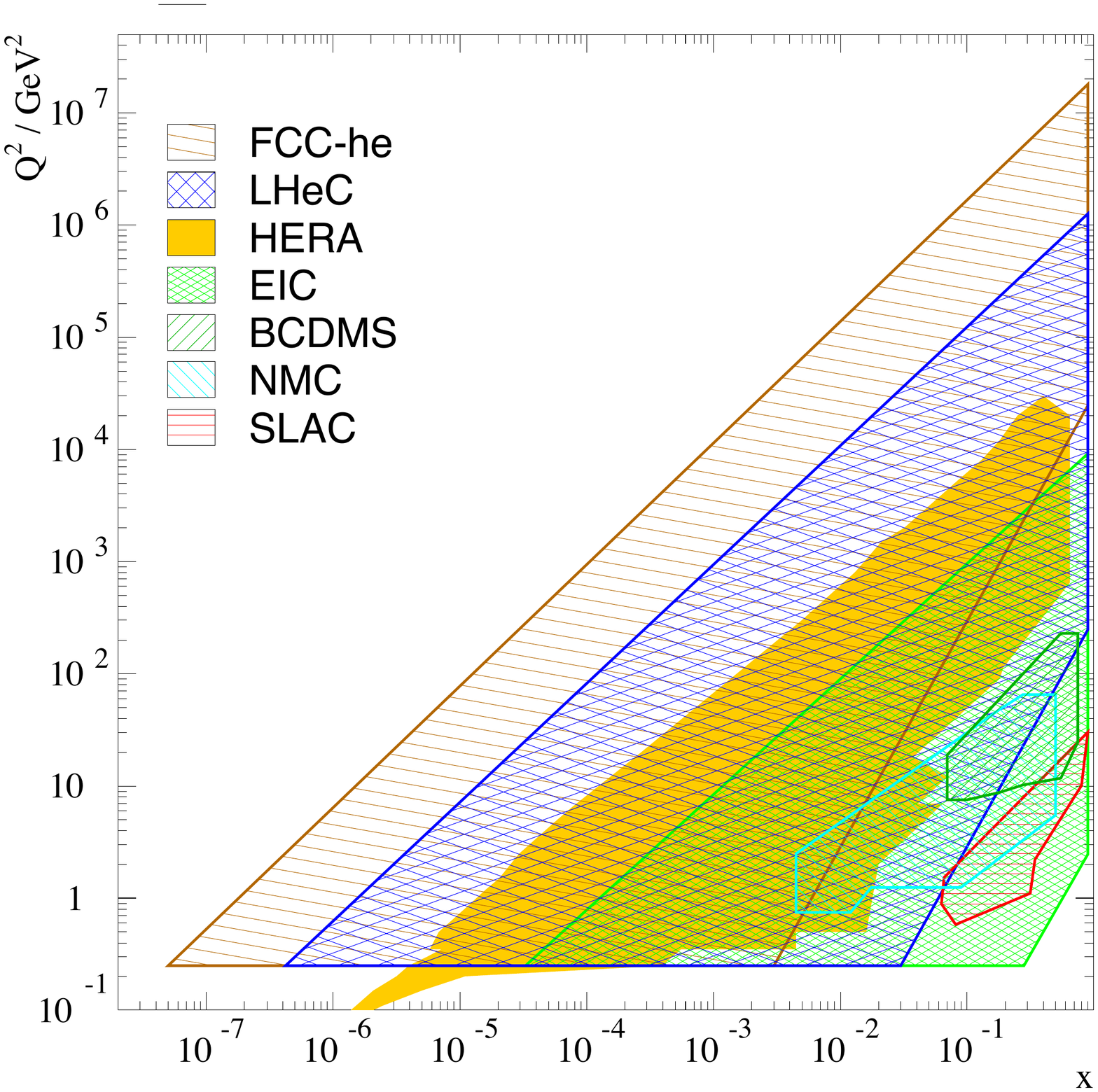} \ \
\raisebox{0.95cm}{\includegraphics[width=0.46\textwidth]{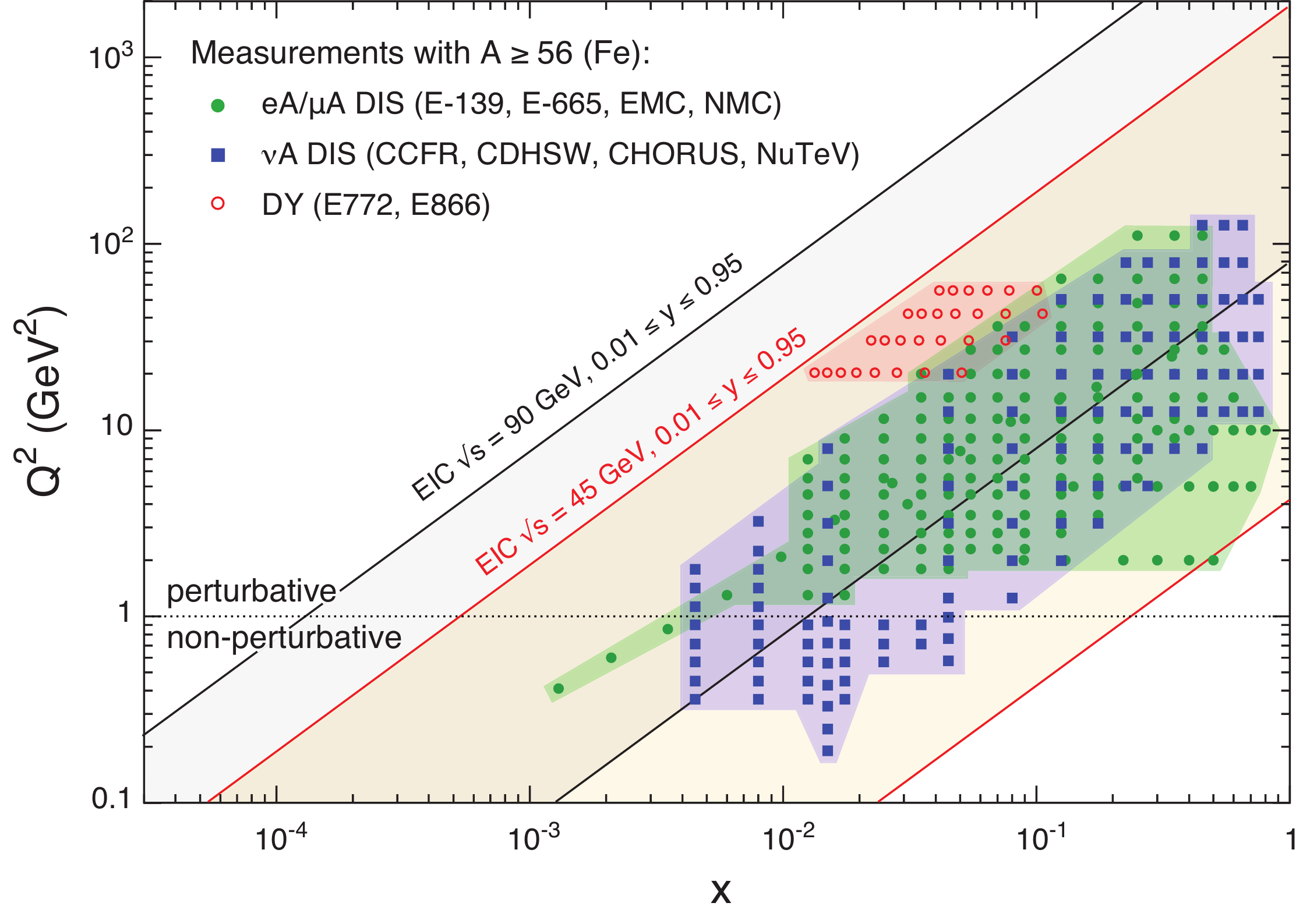}}
\vspace*{-1.05cm}
\caption{(Left) The kinematical coverage of various legacy and upcoming/proposed DIS experiments
	in $x$ and $Q^2$, from Ref.~\cite{LHeC:2020van}. The wide kinematical coverage of
	both the EIC and LHeC intersect the regions probed by older fixed-target DIS experiments
	while extending into novel regions of low $x$ and large $Q^2$. (Right) A complementary
	map of the kinematical coverage of the EIC in DIS measurements involving heavier
	nuclei ($A\! \ge\! 56$); superposed are the placements of legacy data sets involving
	charged-lepton, $\nu$A, and Drell-Yan (DY) information; taken from Ref.~\cite{Accardi:2012qut}.
	}
\label{fig:kinematics}
\end{figure}

In these proceedings, we present a brief overview of two future DIS programs: the US-based Electron-Ion Collider (EIC)~\cite{AbdulKhalek:2021gbh,Accardi:2012qut}
and Large Hadron-electron Collider (LHeC)~\cite{LHeC:2020van} proposed for construction at CERN. In particular, we illustrate how
these programs can be expected to significantly impact precision activities at the LHC and future High-Luminosity LHC
(HL-LHC).
A valuable aspect of the future DIS programs to be carried out at the EIC and LHeC is their ability to probe
complementary regions of the kinematical $(x,Q^2)$ plane as shown in Fig.~\ref{fig:kinematics}.
In particular, the energy and expected luminosity of the EIC program, which
we discuss in greater detail in Sec.~\ref{sec:EIC}, is such
that its primary focus spans the few-GeV (non)perturbative boundary region at very high $x$ and low $Q^2$,
but with robust reach down to $x\! \sim\! 10^{-4}$ and $Q^2\! \sim\! 1000$ GeV$^2$ at larger $x$. The ability
of the EIC to unravel dynamics in the few-GeV transition region to perturbative QCD interactions also applies
to studies involving nuclei, as illustrated in Fig.~\ref{fig:kinematics} (right).
The LHeC discussed in Sec.~\ref{sec:LHeC}, in contrast, will have the ability to probe very low $x\!\sim\! 10^{-6}$ over a significant range of $Q^2$ momenta,
and will be capable of probing perturbative scales as high as $Q^2\! \sim\! 10^{5-6}$ GeV$^2$ by merit of its TeV-regime
kinematics. In consequence of this wide reach, the LHeC would be capable of interrogating overlapping kinematical regions covered
by the HL-LHC, but through measurements of complementary DIS processes. We highlight specific operational parameters
responsible for the unique scope of the EIC and LHeC programs in respective, dedicated subsections below.

We also emphasize that the issues mentioned in these brief proceedings are a small but representative sub-sample of the
numerous points extensively canvassed in the main community literature of the two facilities discussed here:
for the EIC, the recent Yellow Report of Ref.~\cite{AbdulKhalek:2021gbh}; and for the LHeC, the similarly recent
whitepaper, Ref.~\cite{LHeC:2020van}. We refer interested readers to these documents, which have dedicated
sections related to many of the examples noted here.

\begin{figure}[h]
\centering
\vspace*{-0.5cm}
\raisebox{0.8cm}{\includegraphics[width=0.5\textwidth]{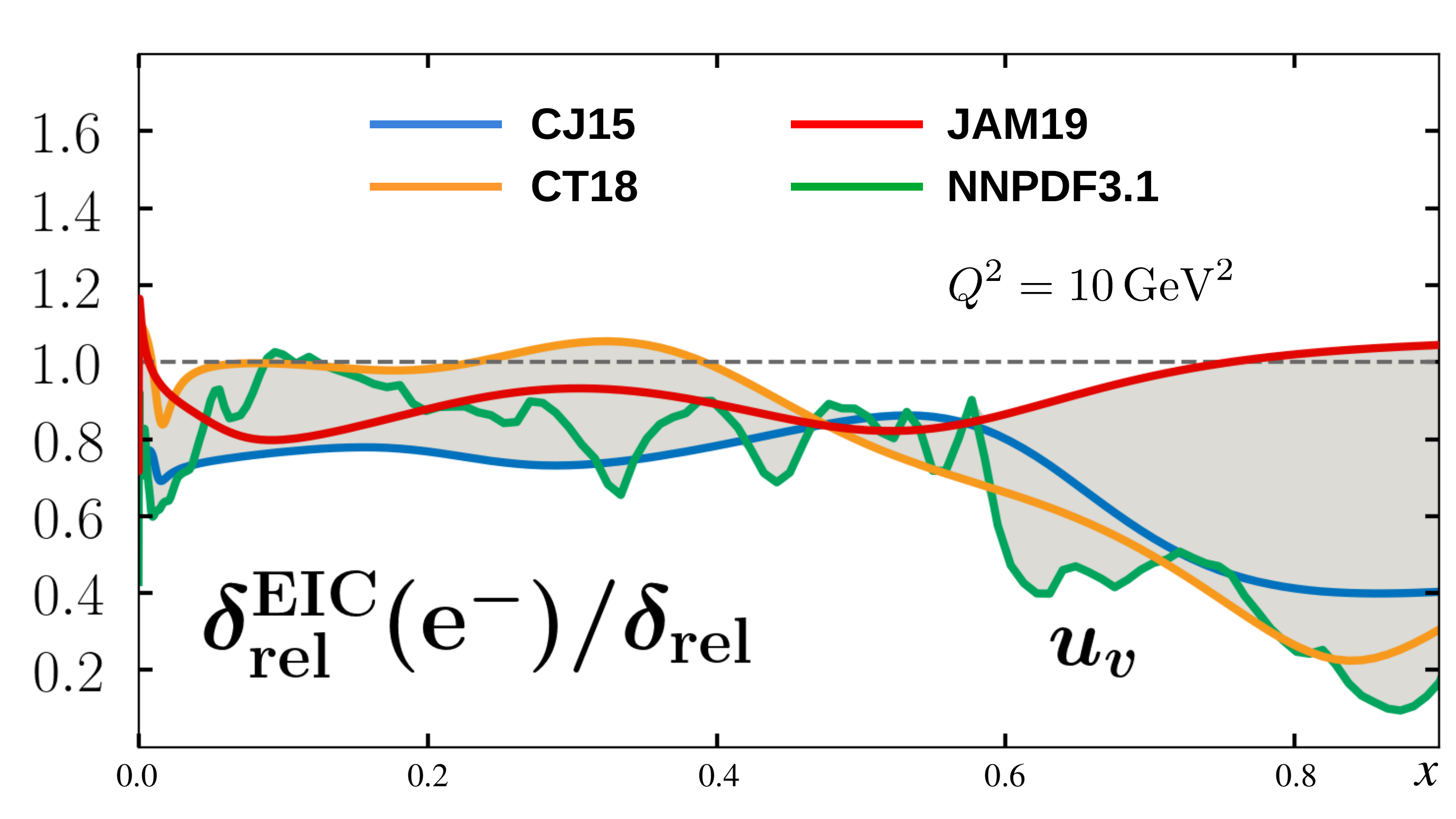}} \ \
\includegraphics[width=0.4\textwidth]{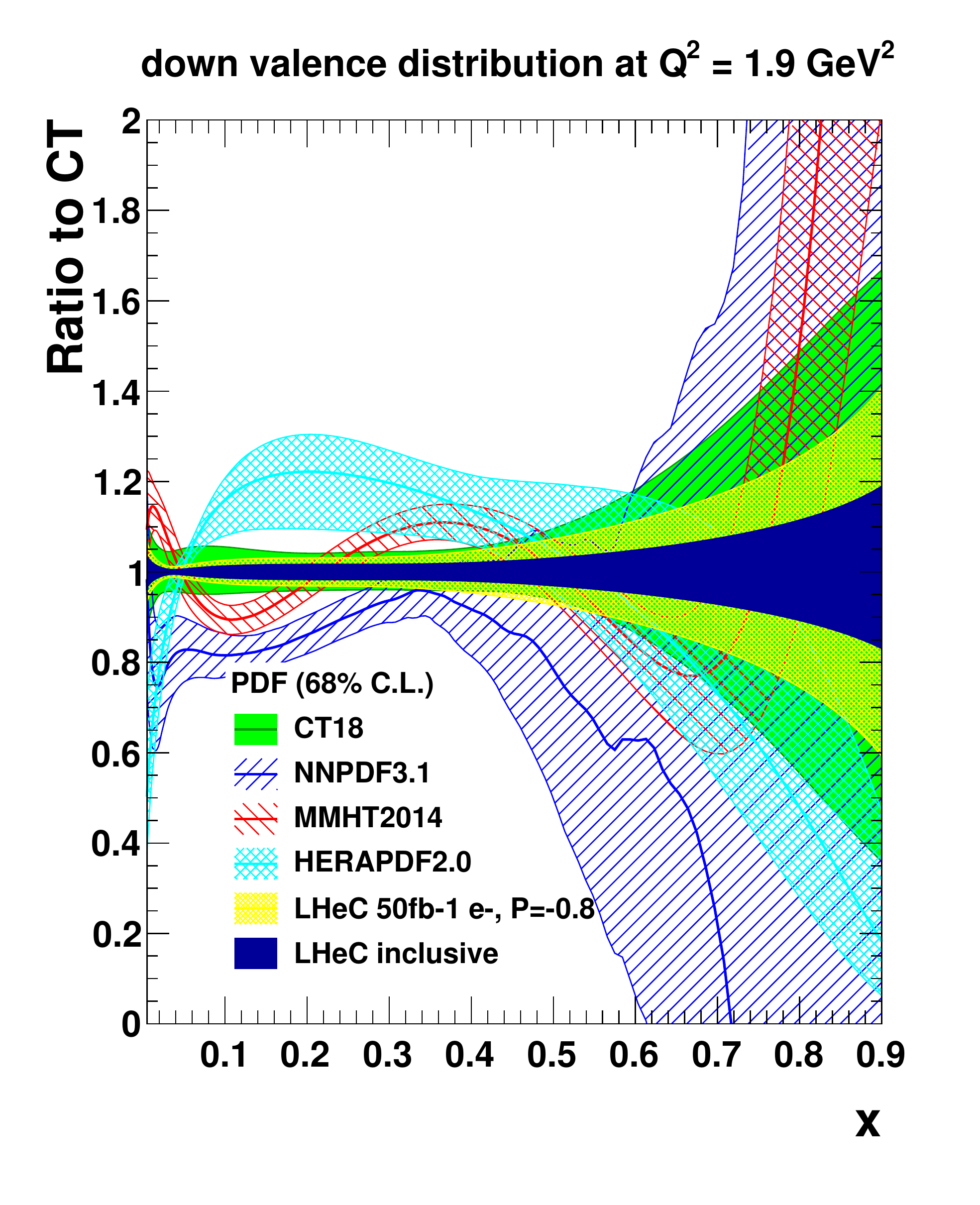} \ \
\vspace*{-0.5cm}
	\caption{
		The EIC and LHeC will have significant constraining power of the unpolarized PDFs of the
		nucleon, as illustrated here for select examples. (Left) The EIC will have the potential
		to reduce high-$x$ uncertainties of the $u_v$ PDF; adapted from Ref.~\cite{AbdulKhalek:2021gbh},
		we plot the relative uncertainty on $u_v(x)$ at $Q^2 = 10$ GeV$^2$ after including $100$
		fb$^{-1}$ of $e^-$ EIC pseudodata, which can reduce PDF uncertainties by as much a factor of
		$\sim\!5$ at $x\sim 0.8$. (Right) Analogously, analyses carried out using the xFitter framework \cite{Alekhin:2014irh}
		suggest that LHeC inclusive measurements can cut down the PDF uncertainty on the $d_v$
		PDF by several factors at high $x$. Such reductions in high-$x$ valence-PDF uncertainties will be
		instrumental for enhancing the precision of BSM searches in, {\it e.g.}, the tails
		of rapidity distributions in high-mass Drell-Yan at the HL-LHC. Panel taken
		from Ref.~\cite{LHeC:2020van}.
	}
\label{fig:PDFs}
\end{figure}

\section{High-energy reach of the EIC}
\label{sec:EIC}

\subsection{EIC brief review}
We quickly summarize some of the specifics of the upcoming EIC program~\cite{AbdulKhalek:2021gbh} from which its unique capabilities
for QCD and hadronic physics are derived.
Having recently received CD-1 approval from the DOE for development at Brookhaven's RHIC facility, the EIC will be a next-generation
DIS collider and the effective successor to the impactful HERA program at DESY,
with significantly greater instantaneous luminosity (by a factor of $10^{2-3}$). This enhanced luminosity is expected to produce
an expansive set of DIS data, of magnitude $\int dt \mathcal{L} \sim 1$ ab$^{-1}$.
The EIC will collide electrons with a variety of nuclear targets, including the proton, deuteron, and $^3$He; electron-nuclear
collisions involving, {\it e.g.}, uranium, will allow a broad program for charged-lepton nuclear DIS.
For $ep$ scattering, the EIC will be capable of collisions with $E_e \le 18$ GeV and $E_p \le 275$ GeV, offering substantial
kinematical coverage in center-of-mass energy, $20 \le \sqrt{s} \le 140$ GeV.
Although not generally included in baseline scenarios in the recent Yellow Report~\cite{AbdulKhalek:2021gbh}, possible
facility upgrades may allow analogous studies using positron beams, which could open a number of channels for
explorations of charge-symmetry violation~\cite{Hobbs:2011vy} in the deuteron system as well as BSM physics.
Critically, the EIC will supply electrons with up to 80\% beam polarization for collisions with unpolarized light and heavy
nuclei. In addition, scattering with polarized proton and light-nuclear beams will also be available for thorough dissections
of the spin structure of the nucleon.

\begin{figure}[h]
\centering
\includegraphics[width=0.48\textwidth]{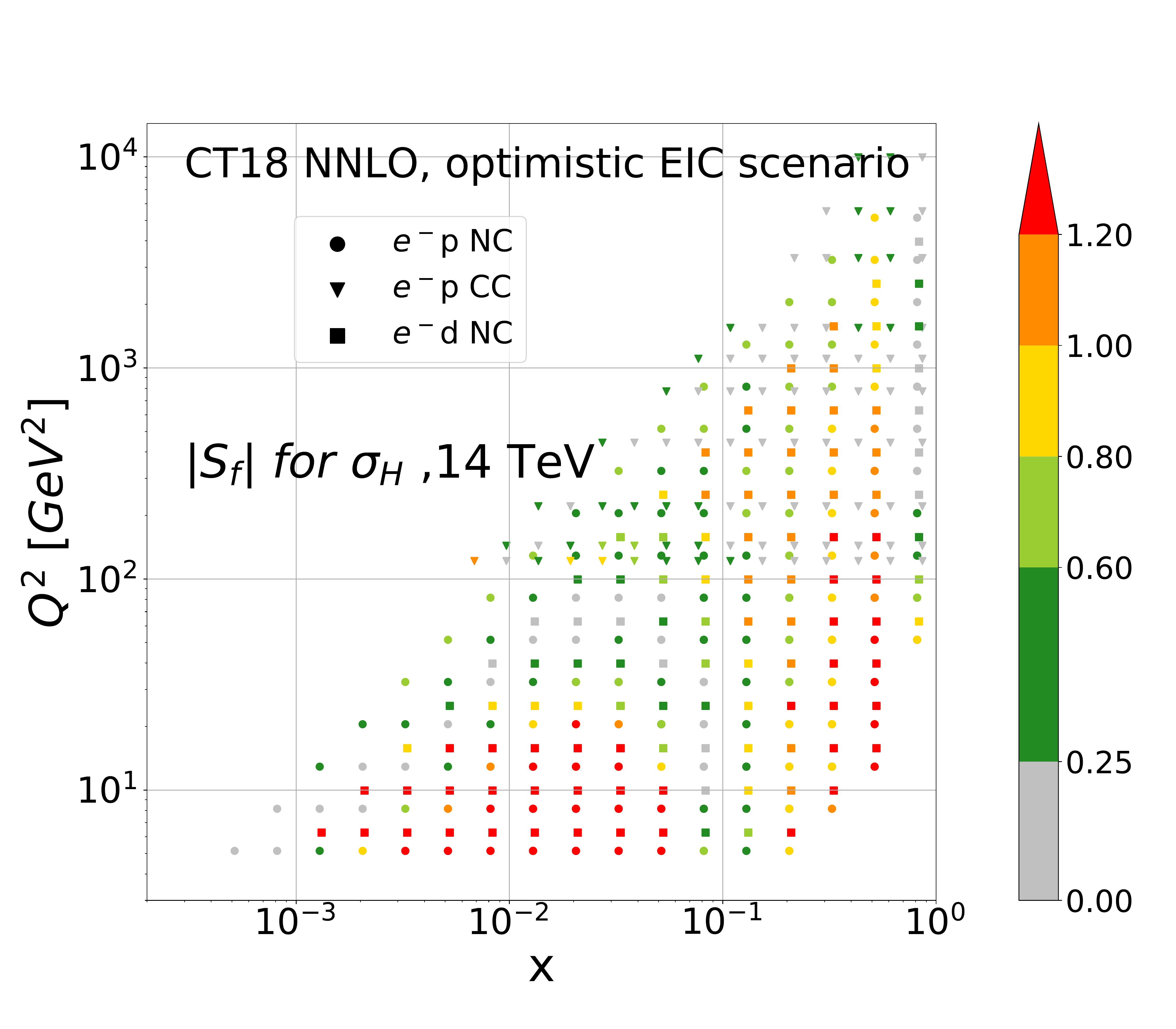} \ \
	\raisebox{0.1cm}{\includegraphics[width=0.46\textwidth]{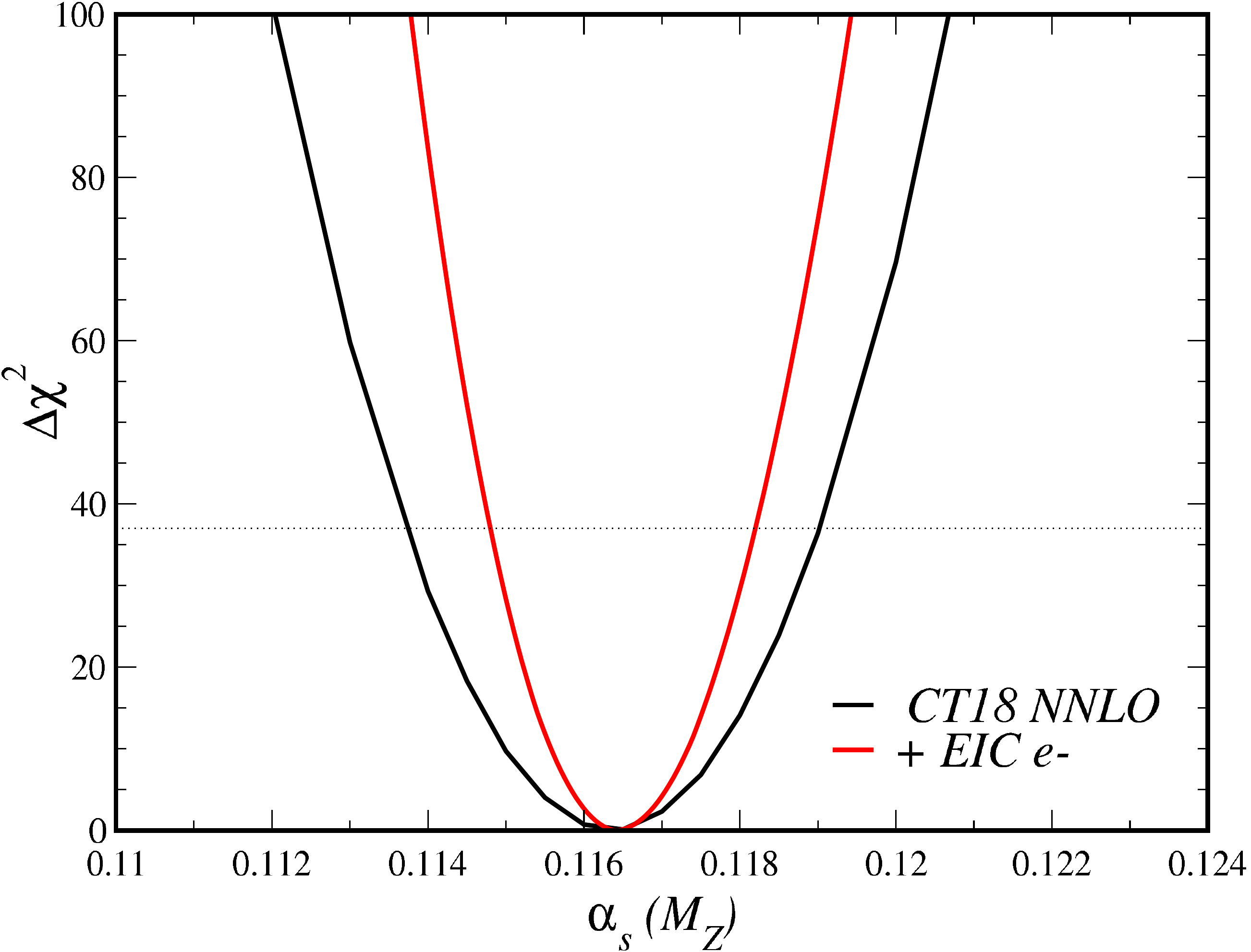}}
\caption{The PDF sensitivity of the $100$ fb$^{-1}$ EIC pseudodata explored in the EIC Yellow Report translates into a
	substantial point-by-point impact on the SM Higgs production cross section (left), as visualized using the
	\texttt{PDFSense} methodology~\cite{Wang:2018heo}. Similarly, extensive DIS collisions covering a range of scales
	and $x$ values results in important potential constraints on the strong coupling, $\alpha_s$ (right); the precision
	on $\alpha_s$ can improve by $\sim 40\%$ over the corresponding uncertainty in CT18 NNLO following the inclusion
	and analysis of $100$ fb$^{-1}$ of electron-scattering data. Both panels are taken from Ref.~\cite{AbdulKhalek:2021gbh}.
	}
\label{fig:EIC}
\end{figure}

\subsection{Tomography implications of the EIC for HEP}
The EIC is a machine chiefly targeted at understanding (non)perturbative QCD and its implications for the
properties of hadrons (including the light mesons~\cite{Arrington:2021biu}) and nuclei, encompassing the multi-dimensional or {\it tomographic}
structure of these strongly-bound systems. To realize these objectives, the EIC program will
consist of an extensive agglomeration of measurements of DIS cross sections and observables of varying inclusivity which will constrain PDFs, TMDs, GPDs,
and hadronic and nuclear form factors --- directly impacting the precision limitations of LHC and $\nu$A measurements that
depend on knowledge of these quantities. The EIC will therefore be capable of sharply resolving the unpolarized PDFs of the proton,
including the valence PDFs like $u_v(x,Q)$ for which we plot in Fig.~\ref{fig:PDFs} (left) the EIC-driven uncertainty reductions recently calculated
for the EIC Yellow Report, Ref.~\cite{AbdulKhalek:2021gbh}; improvements will also extend to the gluon content and flavor structure of the light-quark sea of
importance to Intensity Frontier work in the EW sector. In
this respect, the EIC will be a valuable follow-up to fixed-target Drell-Yan measurements like the recent FNAL SeaQuest (E906)
experiment~\cite{SeaQuest:2021zxb}. At this meeting, we presented some preliminary PDF analysis results, discussed in greater detail in
Ref.~\cite{Guzzi:2021fre}, of the new SeaQuest $\sigma^{pd}/\sigma^{pp}$ data, along with other recent developments in the CT18 NNLO PDF
framework~\cite{Hou:2019efy}.

Beyond the leading-twist unpolarized PDFs of most immediate relevance to LHC predictions, the EIC in particular
will undertake a DIS program beyond fully-inclusive measurements. This will include a variety of transverse-momentum
dependent (TMD) quantities with potential sensitivity to the TMD PDFs and fragmentation functions of the proton.
Such measurements provide an additional setting to perform further test of factorization theorems of QCD as well
as measurements of TMD PDFs related to $W$-mass determinations \cite{Bacchetta:2018lna}.
In addition to tomographic measurements, the EIC will also have a dedicated program
related to perturbative QCD, with the capability of imposing stringent constraints on the strong coupling, $\alpha_s$, heavy-quark
masses, and EW observables. This will be in conjunction with a proposed program to investigate, {\it e.g.}, DIS
jet production, single-inclusive hadron production, and other processes which test QCD in the perturbative regime and facilitate
studies of the applicability and range of validity of various QCD factorization theorems. In addition, processes like charge-current
DIS jet production~\cite{Arratia:2020azl} may unlock novel channels with especially strong sensitivity to the nucleon's strange
content, which also has significant implications for realizing next-generation precision in the EW sector.
Regarding the EIC's potential tomography-mediated impact on HEP observables, we show in the left panel of Fig.~\ref{fig:EIC} the
PDF sensitivity of EIC pseudodata to the SM Higgs-production cross section at LHC energies (here, $\sqrt{s}=14$ TeV), calculated using the \texttt{PDFSense}
package~\cite{Wang:2018heo}. These findings were developed in the context of the CT18 NNLO PDF set in support
of the EIC Yellow-Report Initiative. The pseudodata appearing in Fig.~\ref{fig:EIC} (left) assume $100$ fb$^{-1}$ of DIS data in the
form of inclusive reduced cross, $\sigma(x,y,Q^2)$, from neutral-current (NC) and charged-current (CC) $e^-p$ scattering and NC $e^-d$
interactions. The information here presumes an ``optimistic'' scenario for the systematic uncertainties. We note that
the pseudodata here are those that drove the PDF improvement plotted in Fig.~\ref{fig:PDFs} (left), but are now mapped
point-by-point to their respective values of $(x,Q^2)$ and scored according to their pull on $\sigma_H(\sqrt{s}=14\, \mathrm{TeV})$;
redder points indicate those data with stronger sensitivity to the inclusive Higgs cross section. We emphasize that
the strong PDF sensitivity to $\sigma_H$ appearing in Fig.~\ref{fig:PDFs} (left) reflects the incisive constraints the EIC will
place on the proton's gluon PDF, which propagate to SM predictions for Higgs production in $pp$ scattering through the
dominant $gg \to H$ channel. Analogous PDF-driven improvements from the EIC can be expected to power enhancements
in extractions of $m_W$, $\sin^2\theta_W$, and other PDF-dependent searches for BSM physics at the HL-LHC.
Measuring inclusive DIS cross sections over a wide range of scales is informative at the level of perturbative QCD in addition
to parton distributions. We show this by plotting the uncertainty on $\alpha_s(M_Z)$ in the right panel of Fig.~\ref{fig:EIC},
which we obtain by refitting the default parametrization and data sets of CT18 NNLO over a series of chosen values of $\alpha_s(M_Z)$
in the presence of the EIC pseudodata described above. The resulting $\Delta \chi^2$ growth profile obtained
as $\alpha_s(M_Z)$ is varied can then be compared to the corresponding series of fits without the EIC pseudodata; this
reveals a $\sim\!40\%$ reduction in the $1\sigma$-uncertainty on $\alpha_s(M_Z)$. Similar exercises involving other QCD-sector
SM inputs like the heavy-quark masses imply a powerful potential for exploring perturbative QCD at the EIC.

\begin{figure}[h]
\centering
	\includegraphics[width=0.64\textwidth]{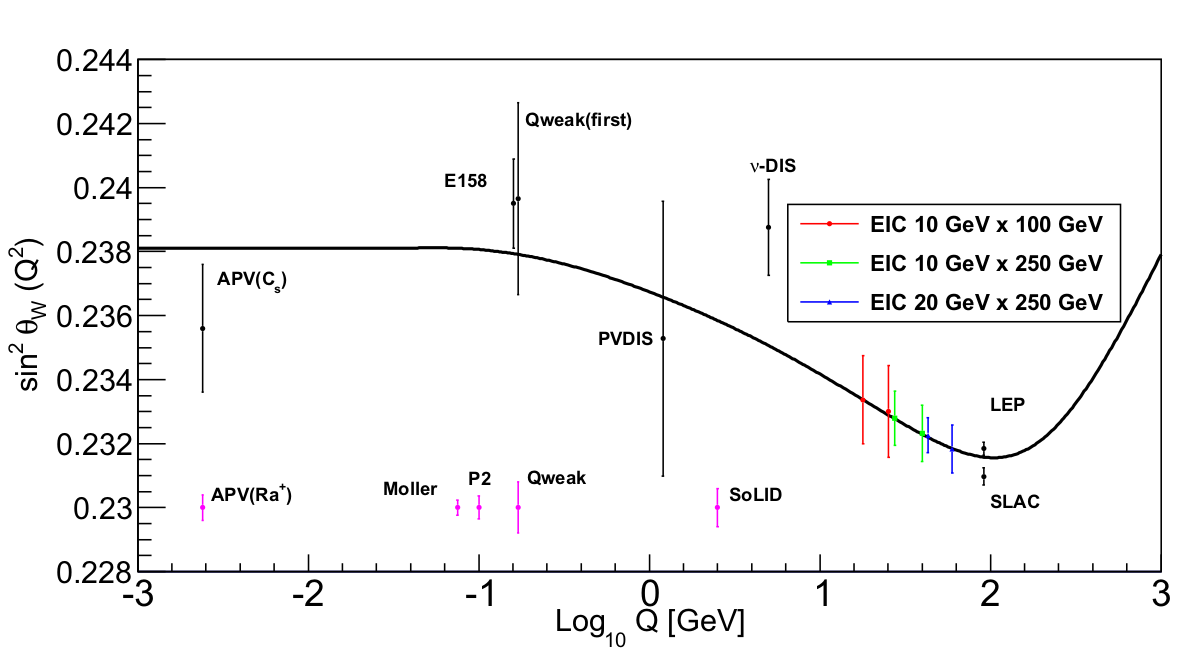}
	\caption{By measuring partiy-violating DIS at the EIC, new coverage and precision can be introduced in determinations
	of $\sin^2\theta_W$, as shown in this plot from Ref.~\cite{Kumar:2016mfi}; this assumes an integrated luminosity
	of $400$ fb$^{-1}$.}
\label{fig:sin2w}
\end{figure}

Finally, we point out that the EIC will have at its disposal the ability to perform a number of more direct probes for
BSM physics, including leptoquark searches for charged-lepton flavor violation (CLFV), as well as precise constraints to
$\sin^2\theta_W$ through parity-violating DIS measurements. In the case of $\sin^2\theta_W$ measurements, the EIC
would have significant constraining power over extensions of the EW sector, including scenarios involving the presence
of hypothetical $Z'$ bosons~\cite{Kumar:2016mfi} as represented in Fig.~\ref{fig:sin2w}. Also noteworthy, {\it polarized} electron-proton DIS
at the EIC has the potential to probe specific Wilson-coefficient combinations otherwise challenging to access purely
through $pp$ scattering data in the context of SM effective field theory (SMEFT) fits as discussed in Ref.~\cite{Boughezal:2020uwq}.
Moreover, the high-precision nuclear physics program at the EIC will similarly entail studies of nuclear-medium effects on parton distributions as well
as DIS jet production from nuclei and explorations of nuclear jet quenching. This work will be informative for $AA$ scattering
at the LHC, and can be expected to benefit, {\it e.g.}, investigations of ultra-peripheral collisions (UPCs).

\section{Precision DIS for HEP at the LHeC}
\label{sec:LHeC}

\subsection{LHeC brief review}
As with the EIC, we provide a quick overview of the specific parameters of the envisioned LHeC facility. Like the EIC,
the LHeC is a proposed DIS machine with extremely high instantaneous luminosity, but also possesses the capability to extend DIS to
the TeV scale for the first time. The LHeC would accomplish this by constructing an Energy Recovery LINAC (ERL) to
provide an electron beam into the HL-LHC complex (or that of the Future Circular Collider, in the FCC-$eh$ proposal).
Kinematically, the LHeC could thus achieve center-of-mass energies of $\sqrt{s}=1.2, 1.3$ TeV by colliding $E_e = 50, 60$ GeV
electrons with protons having the $E_p = 7$ TeV beam energy currently available at the LHC. Strong
electron-beam polarizations with $P_e = \pm 0.8$ would also be possible. By extending DIS to the TeV scale, the LHeC
would offer a sweeping range of channels and possible measurements to examine QCD at high energies --- including
probes of the gluon and quark PDFs like $d_v(x,Q)$ shown in Fig.~\ref{fig:PDFs} (right) to both high and very low
$x$, $x \ge 5 \times 10^{-5}$ --- as well as a compelling battery of SM tests.

\subsection{EW measurements and SM tests at the LHeC}
We note that, like the EIC, extensive DIS measurements at the LHeC would yield a large, $\int dt \mathcal{L} \sim 1-2$ ab$^{-1}$, data set; 
the high-precision constraints from this information to nucleon-level PDFs and related quantities will therefore impose tight bounds on many
PDF-dependent SM quantities, including Higgs and EW cross sections.
Given its access to higher energies, however, the sizes of cross sections for many production processes in the Higgs and EW sectors
will be sufficiently large as to allow precise measurements through DIS production for the first time. In fact, a variety of direct
probes of EW and Higgs physics will therefore be available; as with the EIC-specific discussion in Sec.~\ref{sec:EIC}, we cannot review all
of these, but instead highlight a small number of representative examples.

\begin{figure}[h]
\centering
\includegraphics[width=0.54\textwidth]{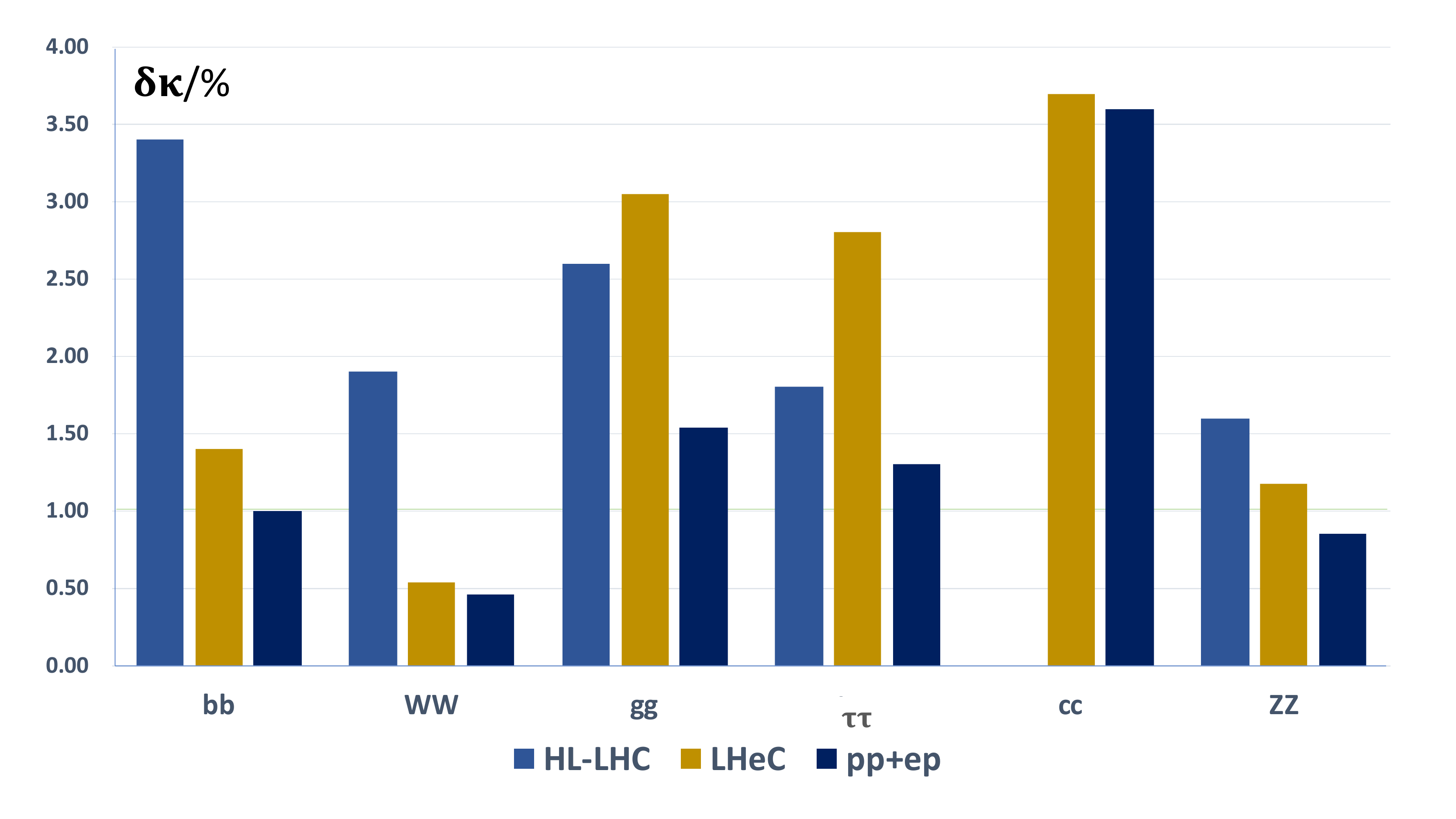} \ \
\includegraphics[width=0.44\textwidth]{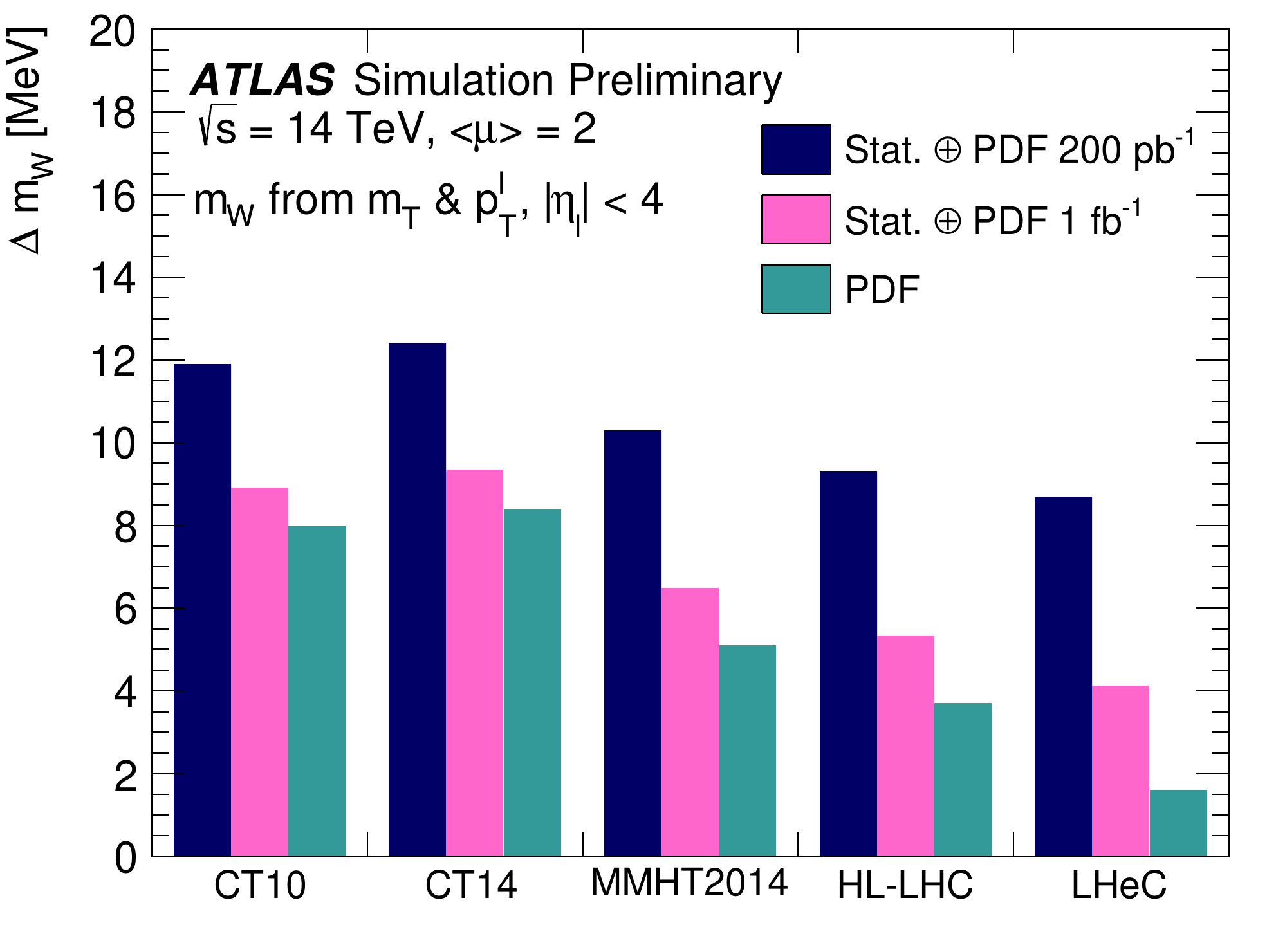}
	\caption{Through direct DIS production, the LHeC will have substantial constraining power over a variety of high-energy observables
	which will impact the HL-LHC program. Inclusion of LHeC pseudodata reduces the uncertainty in determinations of the BSM couplings of the
	Higgs boson (left) to the percent-level or less, particularly for the $bb$, $WW$, and $ZZ$ channels. Similarly, PDF improvements driven
	by the LHeC can dramatically reduce the $W$-mass uncertainty, $\Delta m_W$ (right), by greater than a factor of 2 relative to an HL-LHC-only
	scenario. Both panels are taken from the recent LHeC whitepaper, Ref.~\cite{LHeC:2020van}.}
\label{fig:LHeC}
\end{figure}

In the Higgs sector, the LHeC would substantially complement the reach of the HL-LHC, particularly given the predominance of alternate
production channels in high-energy DIS like $WW \to H$, in contrast to $gg \to H$, which is dominant in $pp$ collisions.
This complementarity can be illustrated quantitatively using the $\kappa_i$ framework~\cite{deBlas:2019rxi}, wherein anomalous
channel-specific deviations from the SM Higgs couplings are parametrized via effective couplings, $\kappa_i$.
The result of a simultaneous analysis of 10 independent $\kappa_i$ couplings was undertaken in Ref.~\cite{LHeC:2020van}, and
we underscore the projected impact from separate HL-LHC and LHeC pseudodata sets in the left panel of Fig.~\ref{fig:LHeC} for
several Higgs decay channels of interest. In addition, the improvement in the uncertainties, $\delta \kappa_i$, achieved
in a combined analysis of $pp$ and $ep$ data is also shown, highlighting the effect of fully leveraging the complementary
hadroproduction and DIS data.
The unique constraining power of the LHeC can be seen in the sharp improvements in the Higgs couplings to $bb$ and $WW$,
for which knowledge approaches the (sub)percent level only after including the high-precision LHeC data set.

The impact of highly constraining data at the LHeC will not be confined to the Higgs sector, but will also influence standard-candle
EW observables, including extractions of $m_W$ as shown in the right panel of Fig.~\ref{fig:LHeC}.
While the upgraded tracking detector at the HL-LHC will afford a greater coverage in the pseudorapidity of the $W$-boson decay
lepton, $|\eta_l| < 4$, extractions of $m_W$ from so-called template fits, in which $m_W$ is tuned in simulations of the kinematic
peaks of final-state distributions, remain PDF dependent. The resulting PDF uncertainty can be as large as $\Delta m_W = \pm 9$ MeV in
contemporary determinations based on ATLAS data \cite{ATLAS:2017rzl}. Extractions based, however, on future PDF sets \cite{ATL-PHYS-PUB-2018-026}
constrained by HL-LHC or LHeC data (the rightmost columns shown in Fig.~\ref{fig:LHeC}), will have dramatically reduced PDF uncertainties,
particularly in the case of LHeC inputs, for which the PDF uncertainty is projected to be as low as $\Delta m_W = \pm 1.6$ MeV.

\section{Conclusion}
The coming decade promises a quickening pace in addressing many of the questions at the heart of QCD with the future
precision DIS programs planned at the EIC and possible at the LHeC. We stress that these programs will leverage a
strong mutual complementarity with activities at the HL-LHC as well as a number of ongoing or planned experiments
from JLab12 to the neutrino DIS programs at DUNE~\cite{1512.06148}, FASER$\nu$~\cite{2001.03073}, and elsewhere. Exploiting this complementarity will
require a continuation of the theoretical improvements necessary to describe hadronic data at a wide range of kinematical
scales and the incorporation of the resulting knowledge into simulations, event-generator calculations, and ongoing
detector design. Development and coordination of these components remains ongoing within the Snowmass 2021 exercises.

\section*{Acknowledgements}
I would like to thank my CTEQ-TEA and nCTEQ colleagues for valuable inputs to this work, with special
thanks to Alberto Accardi, Sayipjamal Dulat, T.-J.~Hou, Xiaoxian Jing, Olek Kusina, Pavel Nadolsky, Fred Olness, Bo-Ting Wang, and C.-P.~Yuan for
providing helpful contributions and suggestions. I also acknowledge numerous valuable recommendations made by members of the EIC
and LHeC communities, including Ludovica Aperio Bella, Abhay Deshpande, Claire Gwenlyn, Douglas Higinbotham, Max Klein, Ute Klein, and Rik Yoshida.
This work was supported by the U.S.~Department of Energy under Grant No.~DE-SC0010129 as well as by a JLab EIC Center
Fellowship. Additional support was provided by the Fermi National Accelerator Laboratory, managed and operated by Fermi Research Alliance, LLC
under Contract No.~DE-AC02-07CH11359 with the U.S.~Department of Energy.

\bibliography{DIS21.bib}

\nolinenumbers

\end{document}